\documentclass{emulateapj}
\usepackage{subfigure}

\shorttitle{Topology of SDSS}

\begin{document}
\title{Topology of Structure in the Sloan Digital Sky Survey: Model Testing} 
\author{J. Richard Gott, III\altaffilmark{1}}
\author{D. Clay Hambrick\altaffilmark{1}}
\author{Michael S. Vogeley\altaffilmark{2}}
\author{Juhan Kim\altaffilmark{3}}
\author{Changbom Park\altaffilmark{3}}
\author{Yun-Young~Choi\altaffilmark{3}}
\author{Renyue Cen\altaffilmark{1}}
\author{Jeremiah P. Ostriker\altaffilmark{1}}
\author{Kentaro Nagamine\altaffilmark{4}}
\altaffiltext{1}{Princeton University Observatory, Princeton, NJ 08544}
\altaffiltext{2}{Department of Physics, Drexel University, Philadelphia, PA 19104}
\altaffiltext{3}{Korea Institute for Advanced Study, CheongNyangNi, Seoul 130-722, Korea}
\altaffiltext{4}{Department of Physics, University of Nevada, Las Vegas, 89154}
\email{dclayh@astro.princeton.edu}

\begin{abstract}
 We measure the three-dimensional topology of large-scale structure in
 the Sloan Digital Sky Survey (SDSS). This allows the genus statistic
 to be measured with unprecedented statistical accuracy.  The sample
 size is now sufficiently large to allow the topology to be an
 important tool for testing galaxy formation models.  For comparison,
 we make mock SDSS samples using several state-of-the-art N-body
 simulations: the Millennium run of Springel et al. (2005)(10 billion
 particles), Kim \& Park (2006) CDM models (1.1 billion particles),
 and Cen \& Ostriker (2006) hydrodynamic code models (8.6 billion
 cell hydro mesh).  Each of these simulations uses a different method
 for modeling galaxy formation.  The SDSS data show a genus curve that is broadly
 characteristic of that produced by Gaussian random phase initial
 conditions.  Thus the data strongly support the standard model of
 inflation where Gaussian random phase initial conditions are produced
 by random quantum fluctuations in the early universe.  But on top of
 this general shape there are measurable differences produced by
 non-linear gravitational effects 
(cf. Matsubara 1994),
 and biasing
 connected with galaxy formation.  The N-body simulations have been
 tuned to reproduce the power spectrum and multiplicity function but
 not topology, so topology is an acid test for these models.  The data
 show a ``meatball'' shift (only partly due to the Sloan Great Wall of
 Galaxies; this shift also appears in a sub-sample not containing the Wall)
 which differs at the $2.5\sigma$ level from the results of the
 Millennium run and the Kim \& Park dark halo models, even including the
 effects of cosmic variance.  

\end{abstract}
\maketitle

\section{Introduction}
The topology of large scale structure in the universe is an important 
physical property of the matter density field that can be compared with
the prediction of the simple inflationary models (Guth 1981; Linde 1983)
where Gaussian random phase initial conditions are generated from 
quantum fluctuations in the early universe. Analytic tools
for quantitatively analyzing the topology of large scale structure 
in three dimensions have been developed during the past 20 years
(Gott, Melott, \& Dickinson 1986;
Hamilton, Gott, \& Weinberg 1986; Gott, Weinberg, \& Melott 1987,
Gott et al. 1989; Vogeley, et al. 1994; Park, Kim, \& Gott 2005;
Park et al. 2005). The distribution of galaxies in space is smoothed
to construct isodensity contour surfaces whose topology may be computed.  
Our genus statistic---described below---quantifies the topology 
of isodensity contours.

On smoothing scales larger than the correlation length, the
fluctuations are still in the linear regime and since fluctuations in
the linear regime grow in place without changing topology, the
topology we measure now should reflect that of the initial conditions,
which should be of Gaussian random phase according to the theory of
inflation \citep{gwm87}.  We have shown this in
detail by comparison with large N-body simulations \citep{gwm87}. 
We expect sponge-like
topology at the median density contour to be a strong prediction of
inflation (Gott, Melott, and Dickinson 1986, Gott, Weinberg, Melott
1987).  Previous models of galaxy clustering suggested either a
meatball topology---isolated clusters growing in a low density
connected background as suggested by the Press and Schechter (1974)
formalism or by hierarchical galaxy formation (Peebles 1974, Soneira
\& Peebles 1978)---or a Swiss cheese topology---isolated voids
surrounded on all sides by walls as suggested by Einasto, Joeveer, \&
Saar (1980).  But with Gaussian random phase initial conditions we
expect a sponge-like topology as pointed out by \citet{gmd86}. 

Studies of many observational samples have been conducted by our group
and others, which have shown in \emph{every} case a sponge-like median
density contour as expected from inflation.  For notable examples, see
Gott, Melott, \& Dickinson (1986), Gott et al.\ (1989), Moore et
al.\ (1992), Vogeley et al.\ (1994), Canaveses et al.\ (1998), Hikage et
al.\ (2002), Hikage et al.\ (2003), and Park et al.\ (2005).  In
addition, in all cases the observed genus curve was reasonably
well-fit by the Gaussian random phase theoretical curve in
equation~\ref{eq:GRgenus}.  Perhaps the most spectacular such
agreement was seen in the Caneveses et al.\ (1998) analysis of the
15,000 galaxy PSCz redshift survey.
This study showed quite a good fit
(within the noise) to the random phase curve at a variety of smoothing
lengths. The IRAS galaxies in this sample are primarily low mass spiral and
irregular galaxies and so may suffer less biasing effects than
galaxies from an optically selected sample. See \citet{park06} for
discussion of the strong dependence of 
morphological fraction on density in the SDSS.

A two-dimensional variant of the genus statistic can also be applied
to redshift slices, sky maps, and CMB maps (Coles 1988, Melott et
al.\ 1989, Gott et al.\ 1990, Park et al.\ 1998), In this case, $G(\nu) =
\#$ of hot (or high density) spots$\mbox{ } - \#$ of cold (or low density)
spots, and the Gaussian random phase hypothesis implies $g(\nu)
\propto \nu \exp(-\nu^2/2)$.  All of these studies (Redshift Slices:
Park et al. 1992; Colley 1997; Hoyle, Vogeley, \& Gott 2002; 
Hoyle et al.\ 2002; Sky Maps: Gott, Mao, Park, \& Lahav
1992; Park, Gott, \& Choi 2001; CMB Maps: Smoot et al.\ 1994; Colley,
Gott, \& Park 1996; Kogut et al.\ 1996; Colley \& Gott 2003; Park 2003; 
Spergel et al. 2006; \citealt{gott06}) indicate consistency with the
Gaussian random 
phase hypothesis.  The CMB maps are a particularly powerful test of
the Gaussian random phase hypothesis because the fluctuations are
still firmly in the linear regime. The dramatic agreement between the
2D CMB results and the Gaussian random phase hypothesis strongly
supports idea of the standard theory of inflation and that the initial
conditions were truly Gaussian random phase.  

The three-dimensional topology data on galaxy clustering are
particularly interesting because they allow us not only to confirm
their general Gaussian random phase nature (on large scales), but also
to test for non-linear processes and bias involved with galaxy
formation. Matsubara (1994) has discussed how the genus curve may be
altered by second-order non-linear gravitational clustering effects,
which can show 
up at small smoothing lengths.  Vogeley et al. (1994)
explicitly measured the diminution of the genus amplitude caused by
these effects.  Park, Kim, \& Gott  (2005) have studied other important 
alterations which can occur by non-linear gravitational evolution, 
redshift space distortion, and biasing associated with galaxy formation.

In this paper we compute genus curves for volume-limited samples of
the largest galaxy redshift survey to date, the Data Release 5 (DR5) of the
SDSS, and for mock samples from state-of-the-art N-body simulations.
Our goal is to examine whether the models for galaxy formation
represented by these simulations are consistent with the
observations. This is a potentially powerful test, because the input
parameters of the flat $\Lambda$CDM model used in these simulations
were determined by fitting to a host of other observations---CMB
anisotropy data, large-scale power spectrum and correlation function
of galaxies, SNeIa luminosity-distance data, cluster abundances and
baryon fraction, etc.---but not topology of large-scale
structure in the galaxy distribution.  As the basic underpinnings of
the model become more secure, we can turn to more precise testing of
models for the physics of galaxy formation.  Likewise, the methods and
parameters for simulating galaxies have been tuned to match other
observations, but not topology. Thus, this comparison provides an
independent test of the model for structure formation.

\section{The Genus and Related Statistics}
The genus is a measure of the topology of the large scale distribution
of galaxies.
We first smooth the point distribution of galaxy positions (we use
only volume-limited samples in the analysis below) with a Gaussian
smoothing ball of radius $\lambda$
\begin{equation}
W(r) = \frac{1}{(2\pi)^{3/2}} e^{-r^2/2\lambda^2},
\end{equation}
where $\lambda$ is chosen to be greater than or equal to the
correlation length.  In this paper we choose $\lambda = 6 h^{-1}$ Mpc
which is approximately equal to the galaxy correlation length.  This
smallest scale yields the highest resolution measure of the
three-dimensional topology and the greatest statistical power because
of the large number of resolution elements.  This scale also gives the
greatest amount of information about non-linear gravitational effects
and biasing involved in galaxy formation.
 
We establish density contour surfaces labeled by $\nu$, where the
volume fraction on the high density side of the density contour
surface is $f$: 
\begin{equation}
f = \frac{1}{\sqrt{2\pi}}\int_\nu^\infty e^{-x^2/2} \,dx.
\end{equation}
The genus as a function of $\nu$ is given by
\begin{equation}
G(\nu) = \mbox{\# of donut holes} - \mbox{\# of isolated regions}
\end{equation}
(Gott, Melott, and Dickinson, 1986).  Thus, an isolated cluster has a
genus of $-1$ by this definition.  We have shown that $G(\nu)$ is also
equal to minus the integral of the Gaussian curvature over the area of
the contour surface divided by $4\pi$, which enables us to measure the
genus with a computer program (CONTOUR 3D) (see Gott, Melott, \&
Dickinson 1986 and Gott, Weinberg, \& Melott 1987).

For a Gaussian random phase density field, the genus per unit volume,
$g(\nu)\equiv G(\nu)/V$, is given by
\begin{equation}
\label{eq:GRgenus}
g(\nu) = A(1 - \nu^2)e^{-\nu^2/2},
\end{equation}
where the amplitude $A = (\langle k^2 \rangle /3)^{3/2}/(2\pi^2)$
depends only on the 
average value of $k^2$ integrated over the smoothed power spectrum
(Hamilton, Gott, \& Weinberg 1986; Adler 1981; Doroshkevich 1970; Gott,
Weinberg, \& Melott 1987).  Thus, the amplitude $A$ [units:
$\mbox{genus}/(h^{-1} \mbox{Mpc})^3$] can tell us about the primordial
power spectrum.  For a Gaussian random field, the median density
contour ($\nu = 0$, $f=50\%$ volume enclosed) exhibits a sponge-like
topology (many holes and no isolated regions); the $f = 7\%$ high
density contour ($\nu = 1.5$) shows isolated clusters, while the $f =
93\%$ density contour ($\nu = -1.5$) is dominated by isolated voids.
We call the curves $G(\nu)$ and $g(\nu)$ (which differ only by a
constant factor for a given sample) the ``genus curves''.  

For the purpose of examining departures of the observed genus curve
from the Gaussian random phase prediction, we parameterize the genus
curve by several derived quantities.  First is the best-fit amplitude,
\begin{equation}
\label{eq:amp}
	A = \mbox{amplitude of the genus curve},
\end{equation}
which we measure by least squares fit of the theoretical random phase
curve to the data, fitting only in the range $-1<\nu<1$.  For the
random phase case, this amplitude is proportional to ($\langle k^2\rangle^{3/2}$ )
of the smoothed power spectrum and so gives information about the
primordial power spectrum. For observations, this amplitude appears lower
because of non-linear clustering and biasing due to
coalescence of structures (Park \& Gott 1991b; Vogeley et al. 1994; 
Canavezes et al. 1998) 

We quantify shifts and deviations of the genus curve from the shape of
the random phase curve using the following three variables.  We
measure horizontal shifts of the central part of the genus curve with
\begin{equation}
\label{eq:deltanu}
\Delta\nu = \frac{\int_{-1}^1 g(\nu)\nu\,d\nu}{\int_{-1}^1
  g_{\mbox{rf}}(\nu)\,d\nu}, 
\end{equation}
where $g_{\mbox{rf}}(\nu)$ is the genus of the random phase
curve following the formula in equation~\ref{eq:GRgenus}, using
the fitted amplitude $A$ above.  The
theoretical curve (Eq.~\ref{eq:GRgenus}) has $\Delta\nu$ = 0.  A
negative value of $\Delta\nu$ is called a ``meatball shift'' as it is
caused by a greater prominence of isolated connected high-density
structures which push the genus curve to the left.  A positive value
of $\Delta\nu$ is called a ``bubble shift'' as it can be caused by a
greater prominence of isolated voids, and might be produced by
isolated explosions (Ostriker and Cowie 1981) as opposed to inflation.
A slight, statistically significant ``meatball shift'' ($\Delta\nu <
0$) was observed first by Gott et al. (1989), who examined the CfA,
Giovanelli \& Haynes, and Tully datasets.  In hindsight, one 
can see a slight ``meatball shift'' in the very first genus curve ever
measured (Gott, Weinberg, Melott 1987) and this ``meatball shift'' was
also seen for brighter galaxies in an analysis of an earlier sample 
of the SDSS (Park et al. 2005). This shift is presumably due to
non-linear galaxy clustering and bias associated with galaxy
formation.  

To quantify departures of the observed genus from the random phase
prediction in the region of the genus curve where isolated voids
should dominate, we 
measure
\begin{equation}
  \label{eq:AV}
A_v = \frac{\int_{-2.2}^{-1.2}g(\nu)\,d\nu}
{\int_{-2.2}^{-1.2}g_{\mbox{rf}}(\nu)\,d\nu},
\end{equation}
where $g_{\mbox{rf}}(\nu)$ is again the genus of the best fit random
phase curve following the formula in equation~\ref{eq:GRgenus}
(see \citealt{pkg05} for an explanation of the choice of range in $\nu$).  As
shown in \citet{pkg05}, a value of $A_v < 1$ can be the
result of biasing in galaxy formation because voids are very empty and
can coalesce into a few larger voids. $A_v$ is sensitive to
the number of isolated voids but the density contour (at $\nu =
-1.7$ for example) is given by the volume fraction, so if $A_v$ is less
than 1, and by definition there is the same volume in the low density
regions being measured, there must therefore be fewer but larger
voids.  Non-linear clustering alone at these scales predicts a value 
of $A_v > 1$ for the power spectrum of the $\Lambda CDM$ model we adopt
(see Figure 1 of Park, Kim, \& Gott 2005), so
observing $A_v < 1$ may be an indication of bias in galaxy formation.

Similar to $A_v$, we measure a quantity 
$A_c$ that characterizes departure
from random phase behavior in the part of the genus curve expected to
be sensitive to the number of isolated high-density regions
(clusters),
\begin{equation}
\label{eq:AC}
A_c = \frac{\int_{1.2}^{2.2}g(\nu)\,d\nu}
{\int_{1.2}^{2.2}g_{\mbox{rf}}(\nu)\,d\nu}.
\end{equation}
A value of $A_c < 1$ may occur because of non-linear clustering, when
clusters collide and merge.  Also if there is a single large connected
structure like the Sloan Great Wall, this can also lower the value of
$A_c$. 
Also, as Park, Kim, \& Gott (2005) have shown, the Matsubara
(1994) formula for second-order gravitational non-linear effects has
the result that $A_v + A_c = 2$ at all scales, so if we observe both
$A_v$ and $A_c$ to be less than 1, (as we find below to be the case)
biased galaxy formation must be involved.

\section{N-Body Simulations of Large-Scale Structure}

Before we confront results of current simulations of the flat
$\Lambda$CDM model with the best observations of the topology of the
galaxy distribution currently available, it is instructive to consider
the remarkable success to date of large N-body simulations in modeling
large-scale structure. It is encouraging that as the volume and
resolution in N-body simulations have grown with the size and quality
of observational data sets, that the agreement has become even more
spectacular---perhaps a sign that we are on the right track with these
models.

Peebles did the first large N-body simulation for cosmology
using 1000 dark matter particles with $\Omega_m = 1$ and Poisson initial
conditions.  It showed clusters like the Coma cluster forming from
random fluctuations by gravitational instability and a reasonable
covariance function.  Aarseth, Gott and Turner (1973) used 4,000
particles with initial conditions that had more power on large scales
than Poisson (index $n = -1$).  They found power law covariance
functions quite like those observed even for models with $\Omega_m <
1$ and $n = -1$ (Gott, Turner, \& Aarseth 1979, Gott \& Turner 1979)
as originally proposed theoretically by Gott \& Rees (1975).  (Indeed,
inflationary flat lambda models popular today have $\Omega_m < 1$ and
more power on large scales than Poisson initial conditions, just as
these early simulations suggested.) They also found voids as large as
those observed.  The largest voids had volumes such that at the mean
density they would have contained as much mass as the Coma type
clusters contained.  This was reasonable from theoretical considerations
of non-linear clustering, considering cluster (Gunn and Gott 1972) and
void (Bertschinger 1985, Fillmore \& Goldreich 1984) formation from
small fluctuations via gravitational instability.  In Gaussian random
phase initial conditions, isolated over- and under-dense regions in the
initial conditions should be equal in mass leading to equal mass great
clusters and empty voids lacking the same amount of mass. They also
found that such $\Omega_m < 1$ models with more power at large scales
than Poisson produced better multiplicity functions than $\Omega_m =
1$ Poisson models (Gott \& Turner 1977; Bhavsar, Gott, \& Aarseth
1981).  

The advent of inflation brought for the first time realistic
theoretical power spectra to input into N-body models. Together,
inflation and Cold Dark Matter specified reasonable initial
conditions. A suite of such simulations by Davis et al. (1985)
provided an impressive match to many aspects of the observed
large-scale structure.  

Just as theory seemed to be converging on the now-disproven ``standard
CDM model,'' the observations provided a shock. De Lapparent, Geller
\& Huchra (1986) found many voids $50 h^{-1}$ Mpc across.  This caused
a number of people to abandon Gaussian random phase initial conditions
and gravitational instability---favoring explosions to produce the
voids instead (Ostriker \& Cowie 1981).  Then Geller \& Huchra (1989)
discovered the CfA Great Wall of galaxies, which surprised everyone.
This result was announced at an IAU conference in Rio de Janeiro.
Many people said that was the end for random phase initial conditions,
for one expected the covariance function to die at a scale of about
$30 h^{-1}$ Mpc and here was a structure that was $150 h^{-1}$ Mpc
long. ``Perhaps it was produced by cosmic string wakes,'' was one comment
made at the time (though not by us).

However, the jump to abandon random phase initial conditions ignored
the fact that no N-body simulations had been done by that time that
were large enough to properly model structures as large as the CfA
Great Wall. When Park (1990) did such simulations using 4 million
particles, simulating such a volume for the first time, the results
showed that such great walls form routinely. In fact, a $20^\circ$
thick slice survey through the simulations was a near perfect visual
match to the Geller and Huchra survey.  These simulations used a
standard peak biasing scheme and included both standard CDM and $\Omega_m =
0.4$ models. Narrow $6^\circ$ thick slices showed prominent large
voids like those in the De Lapparent, Geller \& Huchra slice and great
walls appeared in $20^\circ$ thick slices.  This simulation showed
weak narrow walls and filaments of galaxies inside the voids, as seen
in the CfA data (Park, Gott, Melott, \& Karachentsev 1992).

In similar fashion, N-body dark matter simulations large enough to
mimic the deep 
pencil beam surveys of Broadhurst, Ellis, Koo, \& Szalay (1990) showed
apparently regular spikes (walls) of galaxies just like those observed
(Park \& Gott 1991a).  And large N-body simulations (Park 1991) showed
great attractors just like that observed by Lynden-Bell et al.\ (1988).
Great repulsors are not seen because such peaks in the gravitational
potential occur in the middle of large voids where there are too few
tracer galaxies.  N-body simulations including hydrodynamics have been
successful in modeling the Lyman Alpha forest (Cen, Miralda-Escude,
Ostriker, \& Rauch 1994; Hernquist, Katz, Weinberg, \&
Miralda-Escude 1996).

Prior to this study we did an analysis of the topology of a large
N-body computer simulation made to mimic the Sloan Digital Sky Survey
(Colley et al. 2000).  This 54 million particle simulation was
observed from one location to simulate what will be seen by the Sloan
Digital Sky Survey, to produce sky maps, slices, and 3D topology maps,
to show the power of the survey.  The sky map looked astonishingly
like real sky maps made to similar depth, and the slice maps looked
quite like similar survey maps made in the Las Campanas Survey \citep{kirsh81}
and now seen in the SDSS.  The cosmological model for this simulation
was the flat $\Lambda$CDM model which remains in favor.

Even larger simulations are available today and we are interested to
see how they fare in their ability to model the topology of
large-scale structure.  The ``Millennium Run'' (MR hereafter), using
over 10 billion ($2160^3$) dark matter particles \citep{springel05} and
surveying a cube of side length $500 h^{-1}$ Mpc, has shown structures
remarkably like the Great Wall found by Geller \& Huchra (1989), and
even wall complexes somewhat resembling the Sloan Great Wall which Gott et al.\ (2006) measured to span 1.37
billion light years (Springel, Frenk \& White 2006).  Indeed Figure 1
in \citet{springel06} shows a remarkable visual agreement between what
is seen in the Millennium Run and in slices of the CfA, the 2dF survey
and the SDSS.  The most noticeable difference is that the Sloan Great
Wall looks much more visually prominent and coherent than the
longest chain of walls found in the MR.  (In this simulation, the box
size of $500 h^{-1}$ Mpc cuts off the power spectrum at larger
scales.  If a simulation were to be made with a larger box size, it
would have more power at these larger scales and therefore could more
easily produce large coherent structures like the Sloan Great Wall.)
The MR computes dark matter halo formation 
merger trees and uses a semi-analytic model to simulate the
galaxy-formation process where star formation and feedback are modeled
by simple analytic physical models. 
\citet{croton06} have produced
mock galaxy samples of their cube that include galaxies brighter than
the Magellanic clouds, including absolute
magnitudes on the SDSS system, which allow us to make mock SDSS
galaxy samples.

Park, Kim \& Gott (2005) produced 8.6 billion particle ($2048^3$)
simulations that cover volumes of $(1024 h^{-1} \mbox{Mpc})^3$
and $(5632 h^{-1} \mbox{Mpc})^3$.
These simulations employ a Halo Occupation Distribution (HOD) method
to place an appropriate number of galaxies in heavy halos identified
by the PSB (Kim \& Park 2006) and FoF techniques. These simulations were used to analyze
the effects of galaxy formation and bias on topology by Park, Kim, \&
Gott (2005). More recently, Kim \& Park have produced 1.1
billion particle simulations 
covering a volume of ($614 h^{-1} \mbox{Mpc})^3$.  Here they use a
new technique to identify physically bound dark matter halos (not
tidally disrupted by larger structures) at the present epoch and
identify these with galaxies (we call these the DH simulations, for
Dark (matter) Halos).  They too have produced magnitudes for these
mock galaxies on the SDSS system by matching the halo mass function
with the luminosity function of the SDSS galaxies.

Cen \& Ostriker (2006) have run hydrodynamic simulations covering a
smaller cube of ($120 h^{-1}$ Mpc) on a side using an 8.6 billion
($2048^3$) cell hydro mesh, with ($1024^3$) dark matter particles.
Here the galaxy formation process is simulated 
with a hydrodynamic code that identifies collapsing regions,
calculates star formation rates, and includes radiative cooling/heating,
UV background radiation with local attenuation, and supernova feedback
associated with star formation.  Again, some assumptions about star
formation are made, but this model has one of the most detailed and direct
physical calculations of the galaxy formation process available for any
simulation that spans a cosmologically-interesting volume.   For
further details of the simulation, we refer the 
readers to \citet{cno05}. Nagamine has produced the mock catalogs
giving absolute magnitudes  on the SDSS system from the Cen \& Ostriker
(2006) hydro simulations.

The MR simulation, the Kim \& Park DH simulation and the Cen \&
Ostriker hydro simulation represent state-of-the-art
simulations for different schemes to mimic galaxy formation.  While
the astronomical community seems to be converging on a standard model
for cosmology---the flat $\Lambda$CDM model (see, e.g., Reiss et al.\
1998, Perlmutter et al.\ 1999, de Bernadis et al.\ 2000, and Spergel et
al.\ 2006)---galaxy formation remains an unsolved problem.  This means
that since only one cosmological model need be simulated, larger N-body
runs exploring different galaxy formation scenarios from different
teams can be run. Since the parameters in the semi-analytic models
have been tuned to account for other features such
as covariance function and multiplicity function, and topology was not
considered, topology is a particularly stringent test.  If the models
produce the right topology automatically, it would constitute dramatic
evidence that their galaxy formation scenarios were on the right
track. In any case, a successful model must show the universe in all
its features, including topology.

\section{Sloan Digital Sky Survey Data}

The SDSS (York et al.\ 2000; Stoughton et al.\ 2002; Adelman-McCarthy et
al.\ 2006) is a survey to explore the large scale distribution of
galaxies and quasars by using a dedicated $2.5 {\rm m}$ telescope
(Gunn et al. 2006) at Apache Point Observatory. The photometric survey
has imaged roughly $\pi$ steradians of the Northern Galactic Cap in
five photometric bandpasses denoted by $u$, $g$, $r$, $i$, and $z$
centered at $3551, 4686,6165, 7481,$ and $8931 \AA $, respectively, by
an imaging camera with 54 CCDs (Fukugita et al. 1996; Gunn et
al. 1998). The limiting magnitudes of photometry at a signal-to-noise
ratio of $5:1$ are $22.0, 22.2, 22.2, 21.3$, and $20.5$ in the five
bandpasses, respectively.  The median width of the PSF is 
$1.4\arcsec$, and the photometric uncertainties are $2 \%$ RMS
(Abazajian et al. 2004). See Ivezic et al (2004) for details of
assessment of photometric quality and Tucker et al. (2006) for
discussion of the monitor telescope pipeline employed for calibration.

After image processing (Lupton et al. 2001; Stoughton et al.  2002;
Pier et al. 2003) and calibration (Hogg et al. 2001; Smith et
al. 2002), targets are selected for spectroscopic follow-up
observation. The spectroscopic survey is planned to continue through
2008 as the Legacy survey and yield about $10^6$ galaxy spectra.  The
spectra are obtained by two dual fiber-fed CCD spectrographs.  The
spectral resolution is $\lambda/\Delta \lambda\sim 1,800$, and the RMS
uncertainty in redshift is $\sim 30$ km/s.  Because of the mechanical
constraint of using fibers, no two fibers can be placed closer than
$55\arcsec$ on the same tile. Mainly due to this fiber collision
constraint, incompleteness of the spectroscopy survey reaches about 6\%
(Blanton et al. 2003a) in such a way that regions with high surface
densities of galaxies become less prominent even after adaptive
overlapping of multiple tiles.  This angular variation of sampling
density is accounted for in our analysis.

The SDSS spectroscopy yields three major samples: the main galaxy
sample (Strauss et al. 2002), the luminous red galaxy sample
(Eisenstein et al. 2001), and the quasar sample (Richards et al.
2002).  The main galaxy sample is a magnitude-limited sample with
apparent Petrosian $r$-magnitude cut of
$m_{r,\mathrm{lim}}\approx17.77$ which is the limiting magnitude for
spectroscopy.  It has a further cut in
Petrosian half-light surface brightness
$\mu_{\rm{R50},\mathrm{limit}}=24.5$ mag/arcsec$^2$.  
More details about the survey can be found on the SDSS web site
\footnote{\url{http://www.sdss.org/dr5/}}.

In our study, we use a subsample of SDSS galaxies
known as the New York University Value-Added Galaxy Catalog (NYU-VAGC;
Blanton et al 2005).  This sample is a subset of the recent SDSS Data
Release 5.  One of the products of the NYU-VAGC used here is
Large-Scale Structure sample DR4plus (LSS-DR4plus).  We use galaxies
within the boundaries shown in Figure 1 of \citet{park06}, which
improves the volume-to-surface area ratio of the survey
(important when smoothing). There are also three stripes in the
Southern Galactic Cap observed by 
SDSS. Density estimation is difficult within these narrow stripes, so
we do not use them.   

\begin{figure}
\includegraphics*[width=\linewidth]{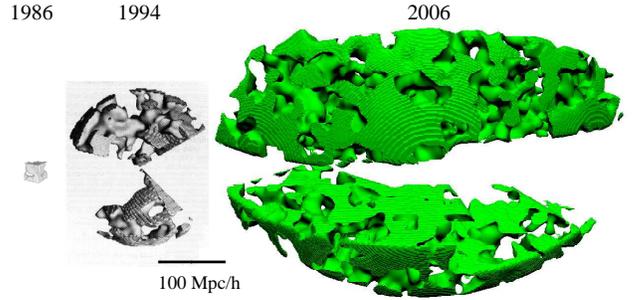}
\caption{50\% high volume contours from three galaxy surveys across
  three decades.  From left to right, they are \citet{gmd86},
  \citet{vogeley94}, and the present work. }
\label{fig:surveys}
\end{figure}

The remaining survey region covers
$4,471$~deg$^2$ ($1.362$ steradians).  The primary sample of galaxies
used here is a subset 
of the LSS-DR4plus sample referred to as {\bf void0}, which is further
selected to have apparent magnitudes in the range $14.5<r<17.6$ and
redshifts in the range $0.001<z<0.5$. These cuts yield a sample of
312,338 galaxies.  The roughly 6\% of targeted galaxies which do not
have a measured redshift due to fiber collisions are assigned the
redshift of their nearest neighbor.
 
Completeness of the SDSS is poor for bright galaxies with $r<14.5$
because of both the spectroscopic selection criteria (which exclude
objects with large flux within the three arcsecond fiber aperture; the
cut at $r=14.5$ is an empirical approximation of the completeness
limit caused by that cut) and the difficulty of obtaining correct
photometry for objects with large angular size.  For these reasons,
analyses of SDSS galaxy samples have typically been limited to
$r>14.5$;  
using the magnitude limits of the {\bf
void0} sample, the range of absolute magnitude is only 3.1 at a
given redshift.

The comoving distance and redshift limits of the volume-limited sample
we analyze are determined from absolute magnitude limits obtained by
using the formula
\begin{equation}
m_{r,{\rm lim}} - M_{r,{\rm lim}} = 5 {\rm log} ((1+z)r) + 25 + \bar{K}(z) + \bar{E}(z),
\end{equation}
where $\bar{K}(z)$ is the mean $K$-correction, $\bar{E}(z)$ is the mean 
luminosity evolution correction,
and $r$ is the comoving distance corresponding to redshift z. 
We adopt a flat $\Lambda$CDM cosmology with density parameters
$\Omega_{\Lambda}=0.73$ and $\Omega_m=0.27$ to convert redshift to 
comoving distance. To determine sample boundaries
we use a polynomial fit to the mean $K$-correction,
\begin{eqnarray}
\bar{K}(z)&=&3.0084(z-0.1){^2}\\\nonumber
&&+1.0543(z-0.1)-2.5 \log (1+0.1).
\end{eqnarray}
We apply the mean luminosity evolution correction given by Tegmark et
al. (2004), $E(z)=1.6(z-0.1)$.  The rest-frame absolute magnitudes of
individual galaxies are computed in fixed bandpasses, shifted to
$z=0.1$, using Galactic reddening corrections (Schlegel 1998) and
$K$-corrections as described by Blanton et al. (2003b).  This means
that a galaxy at $z=0.1$ has a $K$-correction of $-2.5 {\rm
log}(1+0.1)$, independent of its SED.

From this sample, we construct a volume-limited sample containing
galaxies brighter than absolute magnitude $M_r = -20.2$ and fainter than
$M_r = -21.7$, and spanning comoving distance from $171.3 h^{-1}$ Mpc to
$344.5 h^{-1}$ Mpc (corresponding to $z = 0.0578-0.1178$).    
This observational sample is similar to the BEST
sample studied in Park et al. 2005, but now larger in extent.  

This volume-limited sample contains 70,781 galaxies (before overlap
correction) and has a mean galaxy separation of $6.097
h^{-1}$ Mpc, so we can safely apply a Gaussian smoothing length of
$6 h^{-1}$ Mpc.  Numerous numerical experiments have shown that if the
smoothing length is smaller than $1/\sqrt{2} \simeq 0.71$ times the
mean inter-galaxy separation there can be a ``meatball shift'' due to
the algorithm picking out individual galaxies as isolated high density
regions (Gott, Weinberg, \& Melott 1987, Gott et al.\ 1989).  In this
sample the smoothing length is approximately equal to $0.98$ times the
mean interparticle separation so this shot noise effect should be small.
 In any case, this is not critical for our analysis
because we compare the observations directly with mock galaxy catalogs
from N-body simulations.  These mock catalogs are constructed to cover
exactly the same range in absolute magnitude as seen in the
observations and so contain very nearly (within a few percent) the
same total number of galaxies 
in the sample. Because the techniques being applied to the observations
and the N-body simulation mock catalogs are identical, the results
should be identical (within statistical variation) if the N-body
simulations are correctly modeling 
the distribution of galaxies.

\section{Topology of Large-Scale Structure in the SDSS}

In a previous paper (Park et al. 2005) we analyzed the three-dimensional
topology of large-scale structure in the SDSS at a range of smoothing
lengths and compare this with theoretical expectations.  In the
present paper we focus on results with a smoothing length of $6
h^{-1}$ Mpc, which yields the most resolution elements and gives the
most important information on galaxy formation.  The sample of
galaxies available has now grown significantly larger and so we are
now able to make direct comparison of this sample with large N-body
simulations and their various methods of modeling of galaxy formation. As we
shall see, the observational sample is now large enough that the
topology, as measured by the genus curve $g(\nu)$, is now a powerful tool
for testing models of galaxy formation.

Figure~\ref{fig:surveys} shows the progression by date of survey of the 3D
topology of selected galaxy redshift surveys.
All have similar smoothing lengths of $5-6 h^{-1}$ Mpc and show
the median density contour surface with the high density regions shown
as solid and the low density regions as empty.  According to standard
inflationary theory this median density contour should be spongelike.
The small cube on the left shows the 3D region studied by Gott et
al. (1986). The earth is at the lower front right corner of the
cube. The topology is spongelike and the Virgo cluster is included in
the high density region.  The larger region of isodensity contours in
the center of this figure is from the the CfA redshift survey (Vogeley
et al.\ 1994).  The earth is at the center and the upper fan shaped
region is in the North Galactic Hemisphere while the lower fan shaped
region lies within the South Galactic Hemisphere.  The Great Wall
noted by Geller \& Huchra (1989) can be seen connecting high density
regions across the top fan-shaped region.  Again the topology is
spongelike, with the high density regions all connected together, and
the low density regions also connected in an interlocking pattern.

Finally, the portion of the SDSS data now available (in 2006, a full
twenty years after the first figure) is shown on the right of
Figure~\ref{fig:surveys}.  This is the largest region yet studied for
topology and contains nearly 400,000 galaxies in total.  The location of the
earth is at the back.  The horizontal slice extending out toward us is
the northern equatorial slice of the SDSS and includes the
Sloan Great Wall (Gott et al. 2005). The upper slice is a second
contiguous thick region in the northern hemisphere of the SDSS. (When
SDSS-II is complete in 2008, the gap between these two slices
will be filled in.)  It is easy to see that the topology of this
median density contour is spongelike.  The high density regions
(taking up half the volume) form one multiply connected region (shown
solid) and the low density regions (taking up the other half of the
volume) also form one multiply connected region that is interlocking
with the high density region.

\begin{figure}
\includegraphics*[width=\linewidth]{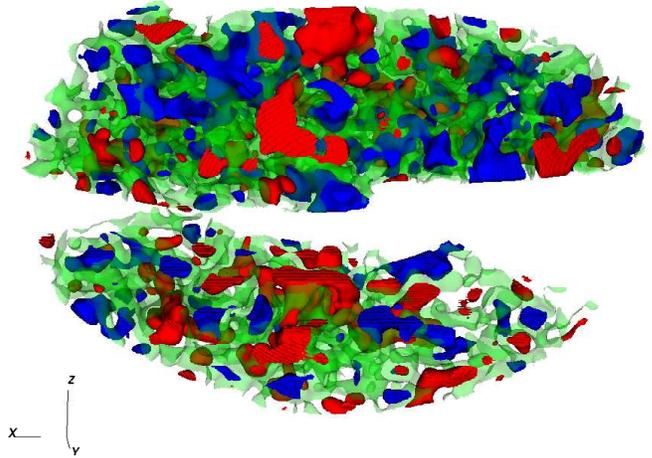}
\caption{7\% low (blue) 50\% (green), and 7\% high (red) volume contours 
 in our SDSS sample. The Sloan Great Wall is visible as the long red
 structure in the lower slice.}
\label{fig:contours}
\end{figure}

Figure~\ref{fig:contours} shows the same regions
of the SDSS, but with isodensity contours in different colors for different
volume fractions.
The 7\% high density
regions---containing the highest density 7\% of the
volume---are solid red.  The end of the Sloan Great Wall can be
seen as the red structure snaking from the left to the right
in the Equatorial slice.  This red contour also shows isolated high density
regions (clusters) as expected from the random phase genus
curve.  The 50\% high density contour is shown in transparent
green---this contour is a multiply connected spongelike surface that
divides the high density half of the sample from the low
density half.  The 7\% low density regions are shown as solid
blue and show isolated voids.  The red and blue regions lie on
opposite sides of the transparent green spongelike surface.

\begin{figure*}
\centerline{\includegraphics*{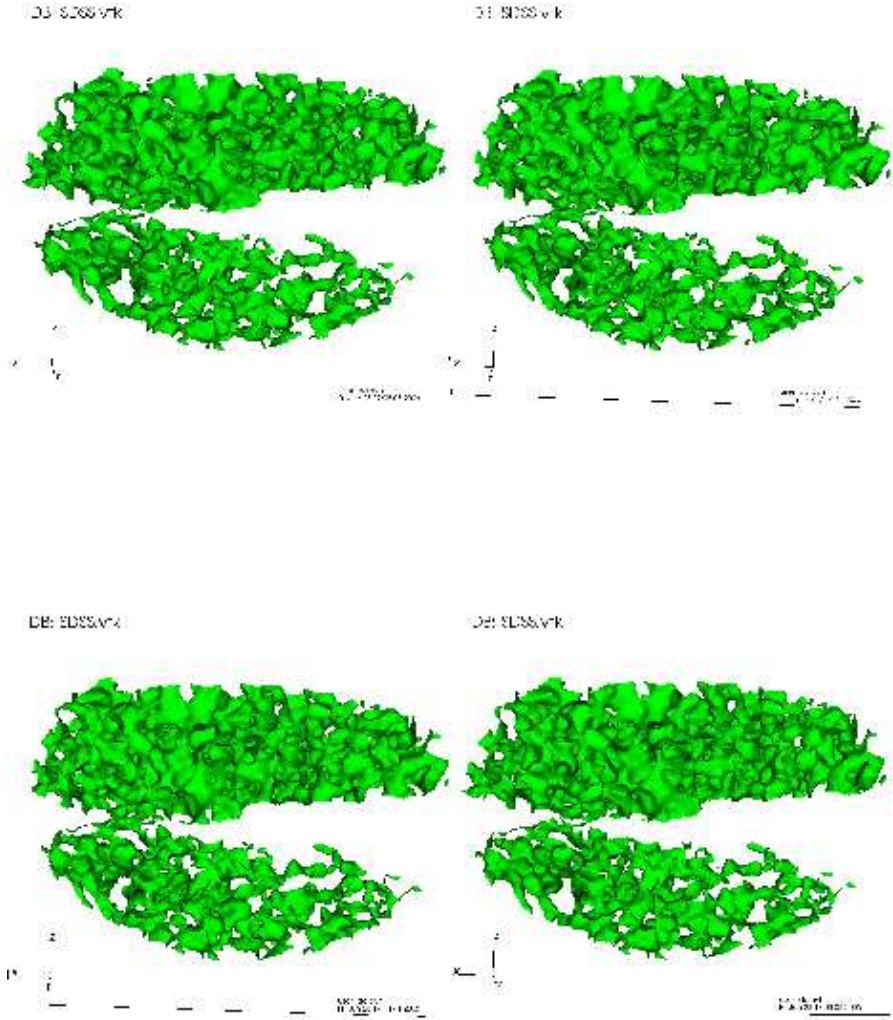}}
\caption{Uncrossed (upper) and crossed (lower) images of the 50\%
 density contour for our sample of the SDSS.  These images are displayed with
separation equal to the separation between your eyes.  To view
the upper pair, touch your nose to the page between the two views.  The
pictures will look blurry but each stereo view will be
directly in front of the proper eye.  Slowly raise your nose
from the paper keeping the blurry images fused into one blurry
stereo view.  As you back away to reading distance you will be
able to bring these two fused views into focus so that you can
see the 3D image clearly.  (A stereo viewer may also be used.)  To see
the lower pair, cross your eyes by 
looking at an object (finger or pen tip) held in front of the page,
moving it until you see three images, then shift your gaze to the
central image.}
\label{fig:stereo}
\end{figure*}

To facilitate viewing the three-dimensional nature of the density contours,
Figure~\ref{fig:stereo} shows a stereo pair of the same SDSS contours.
This offers our best picture yet
of the 3D topology of large scale structure in the universe.   

\begin{figure}
\centerline{\includegraphics[width=\linewidth]{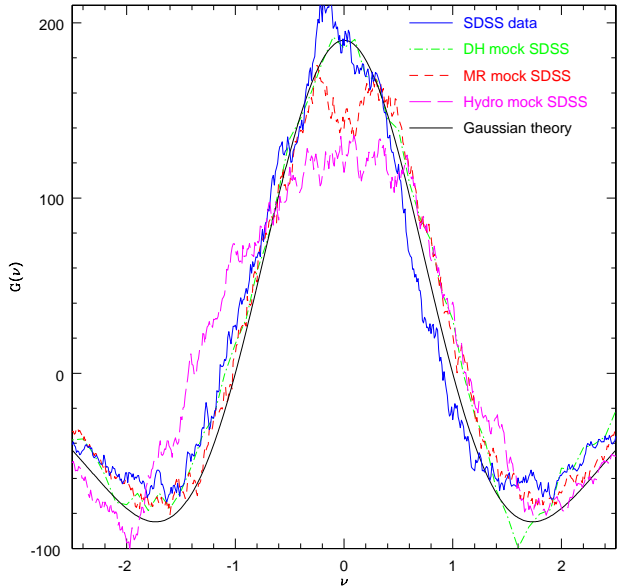}}
\caption{Genus curves for the SDSS sample, hydro sample,  and
 random samples drawn from the 100 DH and 50 MR studies.  The Gaussian
 random curve is shown for comparison.  Notice that we plot $G(\nu)$
 (i.e., do not divide by the sample volume) in order to show the
 genus of the entire sample at each $\nu$.}
\label{fig:genus:r}
\end{figure}

\begin{figure}
\centerline{\includegraphics[width=\linewidth]{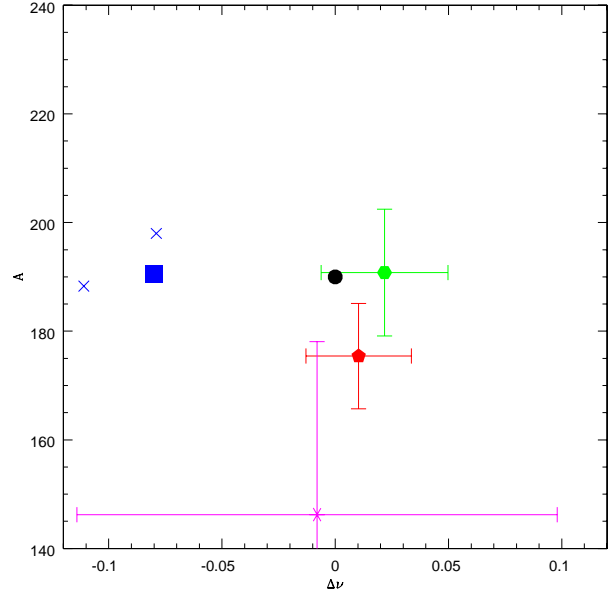}}
\caption{Plot of the genus shift parameter $\Delta\nu$ versus the
  genus-curve amplitude $A$ for our samples.  The black circle
  corresponds to  Gaussian random phase.  The blue square is the
  SDSS sample, with the two blue Xs representing each of the two SDSS
  subregions (normalized in amplitude to the volume of the whole
  sample).  The green 
  hexagon is the mean of the 100 DH mock Sloans, 
  with error bars representing the standard deviation of that sample.
  The red pentagon is the mean of the 50 MR mock Sloans,
  with error bars showing that standard deviation.
  The pink X denotes the hydrodynamic
  simulation of Cen \& Ostriker; the error bars correspond to the cosmic
  variance of ($120 h^{-1}$ Mpc)$^3$ subregions within a ($500 
  h^{-1}$ Mpc)$^3$ box.
  }
\label{fig:ampdn}
\end{figure}

\begin{figure}
\centerline{\includegraphics[width=\linewidth]{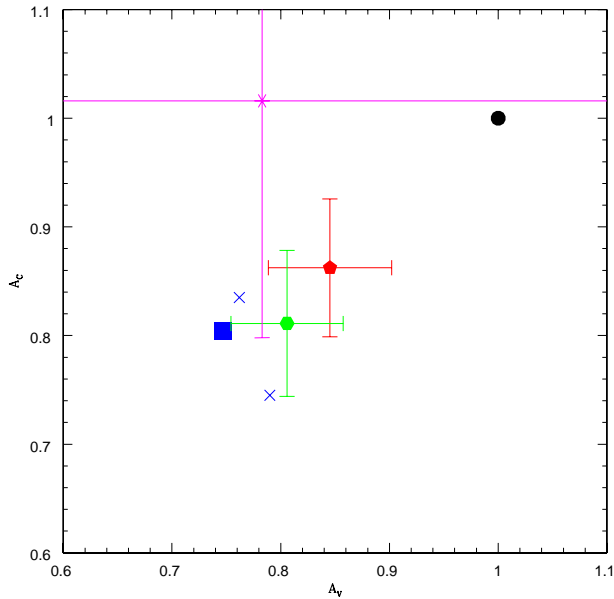}}
\caption{The same as Fig.~\ref{fig:ampdn}, plotting $A_v$ and $A_c$.
  The upper-left blue X corresponds to the SDSS region without the
  Sloan Great Wall.}
\label{fig:AVAC}
\end{figure}

Figure~\ref{fig:genus:r} shows the genus curve of this volume-limited
SDSS sample smoothed at $\lambda=6 h^{-1}$ Mpc. In this figure we
compare the observed genus curve with results for mock surveys
produced from the N-body simulations, which we describe in Section~\ref{sect:sims}
below. Also shown in Figure~\ref{fig:genus:r} is the random phase
genus curve that best fits the SDSS genus curve.  The data
approximately follow this random phase curve, as expected from
inflation. However, there are measurable departures that are likely to
have been caused by non-linear effects and galaxy formation.  We
characterize these differences from the random phase curve using the
measures $A$, $\Delta\nu$, $A_v$, and $A_c$ described above and plot
these values in Figures~\ref{fig:ampdn}~and~\ref{fig:AVAC}.

Figure~\ref{fig:ampdn} shows the $(\Delta\nu, A)$ plane with the SDSS
values of $(\Delta\nu, A)$ plotted as a solid blue square, while the
black circle indicates
the random phase prediction of $\Delta\nu=0$ (the amplitude
of this point is not that of a random phase distribution, but rather
is fixed to be the same as the SDSS data).

The two blue X's show the values computed from the two SDSS slices
separately---as if each were the only sample studied. In this case,
the median density contour is calculated for each sub-sample
separately, thus each has a somewhat different density level at the
median contour because presence of the Sloan Great Wall in the
equatorial sample guarantees that the median density in this sample is
greater than in the northern sample. The difference between the genus
parameters of these two sub-samples provides a rough measure of the
cosmic variance.  
The values of the genus amplitude,
$A$, for the two subsamples are of course smaller, which 
we  correct for 
by dividing each by the respective fraction of the entire
volume it represents.  

The observed genus curve in Figure~\ref{fig:genus:r} is displaced
slightly to the left in the central regions (a slight meatball
shift, or a prominence of clusters over voids).  This gives the curve
a negative value of $\Delta\nu$ as 
indicated by equation~\ref{eq:deltanu}.  The whole sample has a value
of $\Delta\nu = -0.08$.
The SDSS
Great Wall itself is a very prominent simply-connected high density
region and so by itself can cause a meatball shift.
Indeed, the equatorial sample which contains the Sloan Great Wall has
a value of $\Delta\nu = -0.11$. However the northern sample which
does not contain the Sloan Great Wall has by itself a value of
$\Delta\nu = -0.08$.  Thus, the meatball shift in the data is a
general phenomenon and is not due solely to the Sloan Great Wall.

Such a meatball shift has been seen in observational samples before.
It was first noticed and commented on by Gott et al. (1989) who
examined the CfA, Giovanelli \& Haynes, and Tully samples.  Gott, Cen,
and Ostriker (1996) found that hydrodynamic simulations predicted a
meatball shift for (early type) elliptical galaxies relative
to (late type) spiral galaxies (elliptical galaxies tend to congregate
more in isolated rich clusters), an effect later observed in the 2D
topology analysis of the SDSS by Hoyle et al.(2002) where a meatball
shift in red (early type) galaxies was seen relative to blue (late
type) galaxies.  Thus, it is clear that galaxy formation processes can
produce meatball shifts relative to the random phase curve, as we
observe with high statistical significance.  The question is whether
our galaxy formation models accurately reproduce this
effect. Below we discuss whether they successfully model this shift in
the genus curve.

Figure~\ref{fig:AVAC} shows the $(A_v, A_c)$ plane with the SDSS data
again shown 
as a solid blue square, with the random phase prediction shown as a
black circle. Again, X's indicate measurements for the two regions of
the SDSS considered separately.  The data have values of $(A_v, A_c)$
that depart from the random phase values of $(1,1)$ due to biased
galaxy formation and non-linear effects.

\begin{deluxetable*}{cllll}
\tablecaption{Genus statistics for SDSS and simulations thereof
\label{tab:comp}}
\tablehead{
\colhead{Name} & \colhead{Amplitude} & \colhead{$\Delta\nu$} &
  \colhead{$A_v$} & \colhead{$A_c$} }
\startdata
SDSS & 190.46  &$-0.080$ & 0.747 &0.804\\
DH & $190.79 \pm 11.65 $&\hspace{6.5pt}$0.022 \pm 0.028$
   &$0.806 \pm 0.052 $&$ 0.811\pm 0.067$\\ 
MR & $175.42 \pm 9.69 (\pm 9.86)$&\hspace{6.5pt}$ 0.010 \pm 0.023 (\pm 0.036)$
   &$ 0.845 \pm 0.057 (\pm 0.081)$&$ 0.862 \pm 0.063 (\pm 0.097)$\\
Hydro &$146.23 [\pm 31.9](\pm 31.9)$&$-0.008 [\pm 0.106](\pm 0.107)  $
      &$0.783 [\pm 0.328](\pm 0.328)$&$1.016 [\pm 0.218](\pm 0.219)$\\
\enddata
\tablecomments{Errors not in parentheses are the standard deviations
  of the population in question (not given for the real SDSS or the
  Hydro simulation, where there is only one sample); errors
  in parentheses additionally 
  include the effect of cosmic variation within a $1024 h^{-1}$ Mpc
  box, and the bracketed errors for Hydro represent cosmic variation
  for its box size within a  $500 h^{-1}$ Mpc box. }
\end{deluxetable*}

\section{Topology of SDSS Data vs. Simulations}\label{sect:sims}

The genus curve for the SDSS sample shows a clear shift towards a
``meatball'' topology and behavior in the void and cluster-dominated
tails that indicate fewer isolated voids and clusters than expected
from either a Gaussian random phase distribution or from perturbation
theory (see Matsubara 1994 and Park et al. 2005). In this section we
examine whether current simulations of large-scale structure reproduce
these features and use the simulations to assess the statistical
significance of these departures from random phase behavior.

Our approach is to construct mock SDSS redshift samples from each of
the simulations that mimic the observational selection effects caused
by the survey geometry, sampling density of structure, and
redshift-space distortions. In the DH and MR cases, we construct many such mock
surveys, smooth and compute the genus curves, and compute the
variables $A$, $\Delta\nu$, $A_v$, and $A_c$ for each. Then we compute
the mean and standard deviations of those statistics, as plotted in
Figures~\ref{fig:ampdn}~and~\ref{fig:AVAC}.

In considering the predictions of the various N-body simulations it is
important to estimate the cosmic variance one may encounter.  We start
with the DH simulation.  This simulation has a volume of ($614
h^{-1}$ Mpc)$^3$, which is a bit over 16 times the volume of the SDSS
sample, allowing roughly that many independent mock surveys.  We
create 100 such surveys of the SDSS in order to fully sample the
structure in the cube with 
the irregular shape of the SDSS; they are clearly not independent, but the
mean and distribution function of the genus statistics are not
affected by this.  We show the
mean and standard deviations of the genus quantities for the Park et
al.\ simulations in Figures~\ref{fig:ampdn}~and~\ref{fig:AVAC} with a
green hexagon and associated uncertainty limits.  It is clear that the SDSS is
more than $3\sigma$ away from the mean value of $\Delta\nu$.  In fact
none of the 100 mock surveys has a value of $\Delta\nu$ as negative as
the observations. It could be argued that the SDSS contains the SDSS
Great Wall, which is so unusual that this region should be excluded
from the analysis. However, this is exactly the purpose of the 100
mock surveys: to examine the range of values we expect in SDSS-sized
surveys.  So the fact that the SDSS is outside the 2 sigma error bars
from the cosmic variance expected shows that the DH simulations are
not successful in predicting the observed meatball shift. Also, recall
that the values of $\Delta\nu$ for both SDSS sub-samples (one not
containing the Sloan Great Wall) are outside these limits.

The DH simulations are perfect in amplitude relative to the data.
Importantly, the DH simulations (as well as the Millennium Run and the
Cen and Ostriker simulations) include the effects of baryon
oscillations which give the correct initial power spectrum.  A similar
DH simulation with 8.6 billion particles but without baryon
oscillations had an amplitude of $209\pm11$, which is
$1.6\sigma$ high relative to the data (191).  This shows that
the topology can detect the presence of the baryon
oscillations by measuring the slope of the power spectrum---simulations 
without them produce an amplitude of the
genus curve that is too high.  The values of $\Delta\nu = +0.020 \pm
0.027$ for those simulations without the baryon oscillations were not
appreciably different from the ones that include the baryon
oscillations. 

For the MR simulation, we have a volume of ($500
h^{-1}$ Mpc)$^3$ which enables us to make roughly 10 independent mock
surveys of the SDSS.  Again, to fully sample the cube, we make 50 mock
surveys at random position and orientation.
 The mean and standard deviation of these
mock surveys are shown in Figures~\ref{fig:ampdn}~and~\ref{fig:AVAC}
as a red pentagon with associated error bars.  This indicates
the cosmic variance seen from SDSS-sized mock surveys drawn from the
($500 h^{-1}$ Mpc)$^3$ survey region. 

However, we must
additionally consider
the effect of cosmic variance on the scale of the 
simulation region: error bars on simulations with smaller box sizes
will be somewhat underestimated with respect to larger (e.g., the MR
w.r.t.\ the DH), and with respect to the real universe (which of course
has infinite box size).  To estimate
this added variance, we take 8 sub-cubes of volume ($512 h^{-1}$
Mpc)$^3$ out of the larger  DH simulation of Kim \&
Park (2007, in prep.) mentioned above, of volume 
($1024 h^{-1}$ Mpc)$^3$, and make redshift maps of them; i.e.,
we give the galaxies their correct $x$ and $y$ coordinates, but put
them at a $z$ coordinate equal to $z + v_z H_0^{-1}$ as if we were
viewing a redshift space map of the survey region from a great
distance. (Note that even with the different power spectrum, the variance
of the genus statistics should still be correct.)  We then compute the 3D
topology for each of the 8 sub-cubes and 
measure the standard deviation of the 8 mean values of the parameters
$(\Delta\nu, A)$ and $(Av, Ac)$.  This cosmic variance of
the 8 sub-cubes is added to the variance of the DH sample population
(i.e., the standard deviations are added in quadrature) to produce an
effective standard deviation  we expect to find due to cosmic variance
for SDSS-sized surveys in the MR if we had a larger ($1024 h^{-1}$
Mpc)$^3$ simulation from which to draw mock catalogs.
  The increase in the standard deviation is only a few percent for the 
amplitude $A$ but $\sim 50\%$ for the other three genus statistics
(see Table~\ref{tab:comp} for both sets of values).  We would have
liked to make a similar set of sub-cubes for the ($614 h^{-1}$
Mpc)$^3$ DH simulation, but of course one cannot draw a statistically
significant number of samples of that size from a ($1024 h^{-1}$
Mpc)$^3$ box without significant overlap.  

 The fact that the MR has a smaller volume also produces a
systematic effect in that it has less power in its power spectrum at
large scales (no power beyond ($500 h^{-1}$ Mpc)$^3$) so we should
expect its amplitude $A$ to be systematically a little large.
Actually, the amplitude is lower than for the DH simulations and is
about $1\sigma$ low relative to the data so apparently the effects of
cosmic variance in a sample this size are more important than the
systematic effect caused by the lack of long-wavelength modes.  

For the Cen and Ostriker hydro simulation we have only one ($120
h^{-1}$ Mpc)$^3$ simulation volume.  This is about one-eighth the
volume of the SDSS region so using its periodic boundary conditions we
simply replicate it to make a volume large enough to encompass the
SDSS.  We then make a mock redshift catalog with the same
absolute magnitude limits.  Because the periodicity kills the large
scale power we expect the simulation to be choppier than the real
universe and therefore have a higher amplitude $A$ than the
observations.  But even more importantly, we expect a large cosmic
variance in a sample only as small as ($120 h^{-1}$ Mpc)$^3$.  To
model this we construct 64 sub cubes each of volume ($120 h^{-1}$
Mpc)$^3$ from the MR simulation (of volume ($500 h^{-1}$ Mpc)$^3$),
and as above we make redshift maps of them (giving galaxies their correct $x$
and $y$ coordinates but put them at a $z$ coordinate equal to $z + v_z
H_0^{-1}$).  Then we compute the 3D topology for each of the 64 sub
cubes and measure the standard deviation of the 64 values of the
parameters $(\Delta\nu, A)$ and $(A_v, A_c)$.  This cosmic variance
from the 64 sub cubes produces the error bars surrounding the one
value from the Cen et al.\ simulation (shown as a magenta X with error
bars.)  We may then add the extra cosmic variance as per the MR data point
to produce the standard deviations corresponding to a volume of ($1024
h^{-1}$ Mpc)$^3$ (again given in Table~\ref{tab:comp}---the extra st.\
dev.\ is $\lesssim 1\%$ for all statistics).  These error
bars are very large; this one 
simulation is not very constraining of the topological properties.
Roughly speaking, the volume of the Cen and Ostriker simulation is an
eighth that of the SDSS so we expect error bars that are roughly
$\sqrt{8} \simeq 2.8$ times as large as for the cosmic variance seen
in the SDSS. This ratio is approximately correct as
Figures~\ref{fig:ampdn}~\&~\ref{fig:AVAC} show.

In Figure~\ref{fig:genus:r} we show the genus curve for the Cen and
Ostriker simulation.  It is not a good fit to the SDSS observations.
The top of the genus curve is chopped off, and it has a meatball shift
in the void region that is not seen in the observations.  In the
central region $-1 < \nu < 1$, where $\Delta\nu$ is measured, the
curve is too fat, but also has a small negative value of $\Delta\nu =
-0.01$.  This is more negative than either the MR or the
DH simulations, but is still not very close to the observed value
of $\Delta\nu = -0.08$.  The amplitude of the Cen and Ostriker
simulation is much lower than the other three data sets.  Overall the
Cen et al.\ simulation does not 
give a good fit to the data; however, the volume of the simulation is
small and the large error bars show that the observations are within 2
sigma in both $\Delta\nu$ and $A$.  (To be fair, other hydro
simulations by Cen and his collaborators---for example, Gott, Cen, \&
Ostriker 1996---have produced good looking genus curves, so this one
may be suffering from a bit of bad luck.)

So, currently, the Cen and Ostriker hydro simulations are too small.
They need to be increased in volume by about a factor of 10 to be
fairly tested against the SDSS.  In 1975 the largest N-body
cosmological simulation had 4,000 particles (Aarseth, Gott, \& Turner
1975).  In 1990 the largest N-body cosmological simulation had 4
million particles (Park 1990).  This led Gott to predict in 1990 that
by 2005 the record would be 4 billion particles.  Springel et al.\
(2005) did even better than this with over 10 billion particles.  An
increase of a factor of 10 every 5 years is just what would be
predicted by Moore's law (a doubling every 18 months).  So given
Moore's law we should expect that Cen and his colleagues will have a
hydro code simulation with volume 10 times larger by 2011.  Then we
can see if it outperforms the MR or the DH simulations.  Just one
additional hydro run with the same parameters would also be
interesting---would it be closer to the observations (suggesting some
bad luck in the current run) or further away?  With the current error
bars, the MR, DH simulations, and the Cen and Ostriker simulations are
all consistent with each other within $2\sigma$.

In Figure~\ref{fig:AVAC} we compare the results for $A_v$ and
$A_c$. Here the observations and the Millennium Run and the DH
simulations are in better agreement.  The observations have values of
$(A_v, A_c) = (0.75, 0.80)$ which depart from those of the random
phase genus curve $(1,1)$.  As we have mentioned, values of $A_v < 1$ on
these scales are not produced by non-linear gravitational clustering
but must be produced by biased galaxy formation (Park, Kim \& Gott
2005).  Here the Millennium run and the DH simulations are both moved
from the random phase value in the direction of the observations.  The
DH simulation does better, being nearly perfect in $A_c$ and just over
one sigma away in $A_v$.  The Millennium Run is within one sigma in
$A_c$ and around $1.7\sigma$ away in $A_v$ (neglecting cosmic
variance).  Again, the Cen and 
Ostriker simulation is further away in $A_c$, but its larger error
bars leave it within 1 sigma in both $A_v$ and $A_c$.

A possible issue with all these simulations is that their initial
conditions assume a value of $\sigma_8=0.9$ at the present epoch
 to be consistent with the
fit for the WMAP first-year data (Spergel et al. 2003), but the WMAP
three-year data prefer a lower value of $\sigma_8=0.76$ (Spergel et
al. 2006). This implies less non-linear growth of clustering up to the
present epoch, and more biasing. We can estimate the possible effect
of using a smaller 
value of $\sigma_8$ by examining the genus statistics at a slightly
earlier epoch in a simulation designed to reach linear $\sigma_8=0.9$
at $z=0$.  For example, at $z=0.5$ (when $\sigma_8=0.76$) in the DH
simulations, we find that dark matter halos with the same density as
the $z=0$ halos have $\Delta\nu=0.015$, versus
$\Delta\nu=0.02$. Compared to the discrepancy of $\Delta\nu$ between
the SDSS data and simulations, this is an infinitesimal effect and we
conclude that this does not appear to be a problem.  We reach similar
conclusions 
about the effect of $\sigma_8$ on $A_v$ and $A_c$.  The small size of
these effects over this range of redshift (and so, over this same
range of $\sigma_8$) can also be seen in Figure 4 of Park et
al.\ (2005).

Another small but interesting effect has to do with halo or galaxy
identification: nearby halos (or galaxies in the hydro simulation) may
get merged together and only counted as one point.  To estimate this
effect, we calculated the genus of SDSS after merging together all
galaxies closer than 100kpc (an overestimate of the actual scale of
the problem).   This does seem to move SDSS slightly closer to the
simulations: the amplitude decreased by $\sim0.5\%$, $\Delta\nu$ increased
from $-0.080$ to $-0.078$, and $A_v$ increased by
$\sim1\%$, while $A_c$ decreased by about the same amount.  
 
 \begin{figure*}
\mbox{\subfigure[DH]{\includegraphics[width=.50\textwidth]{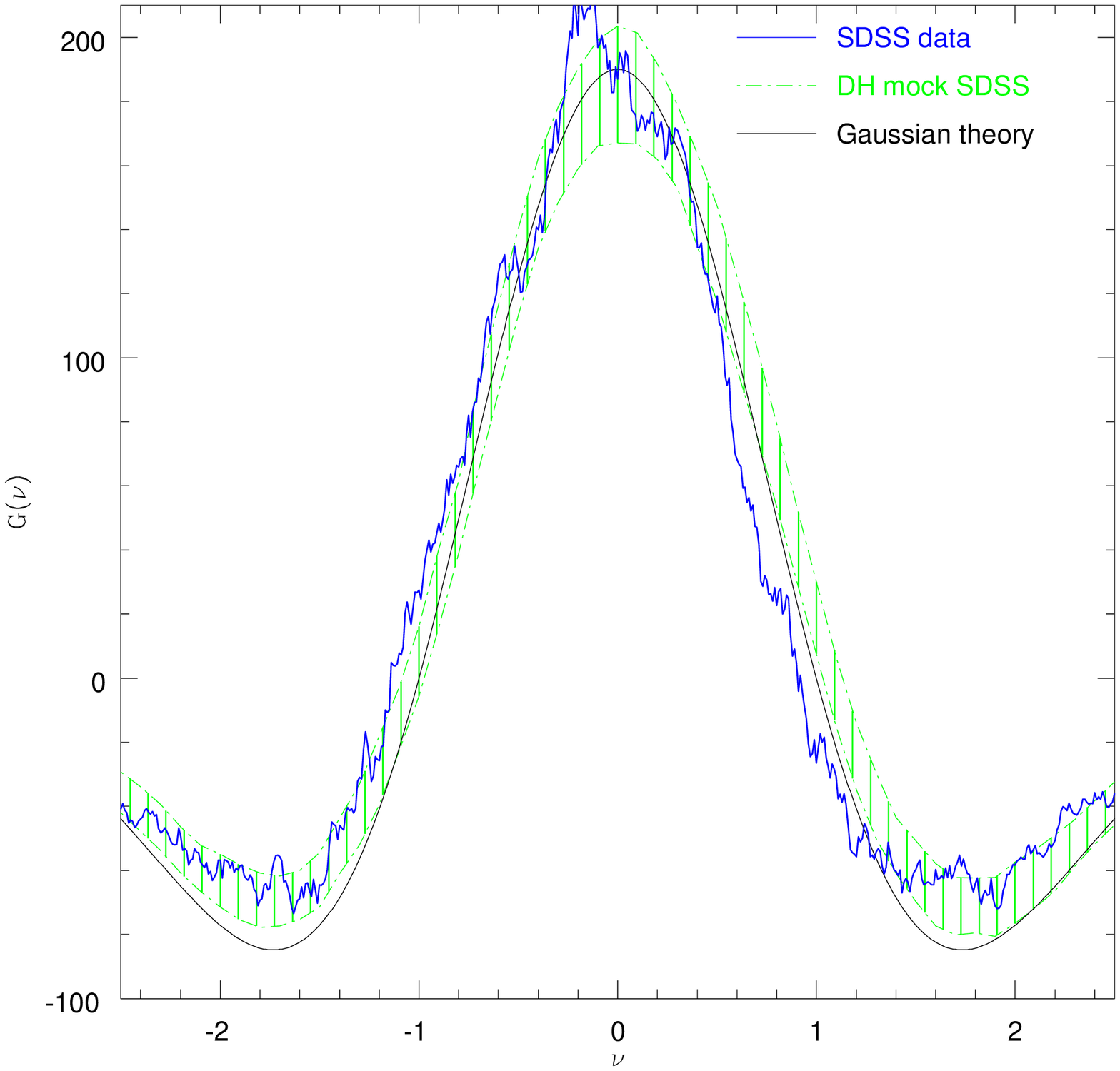}}
\subfigure[MR]{\includegraphics[width=.50\textwidth]{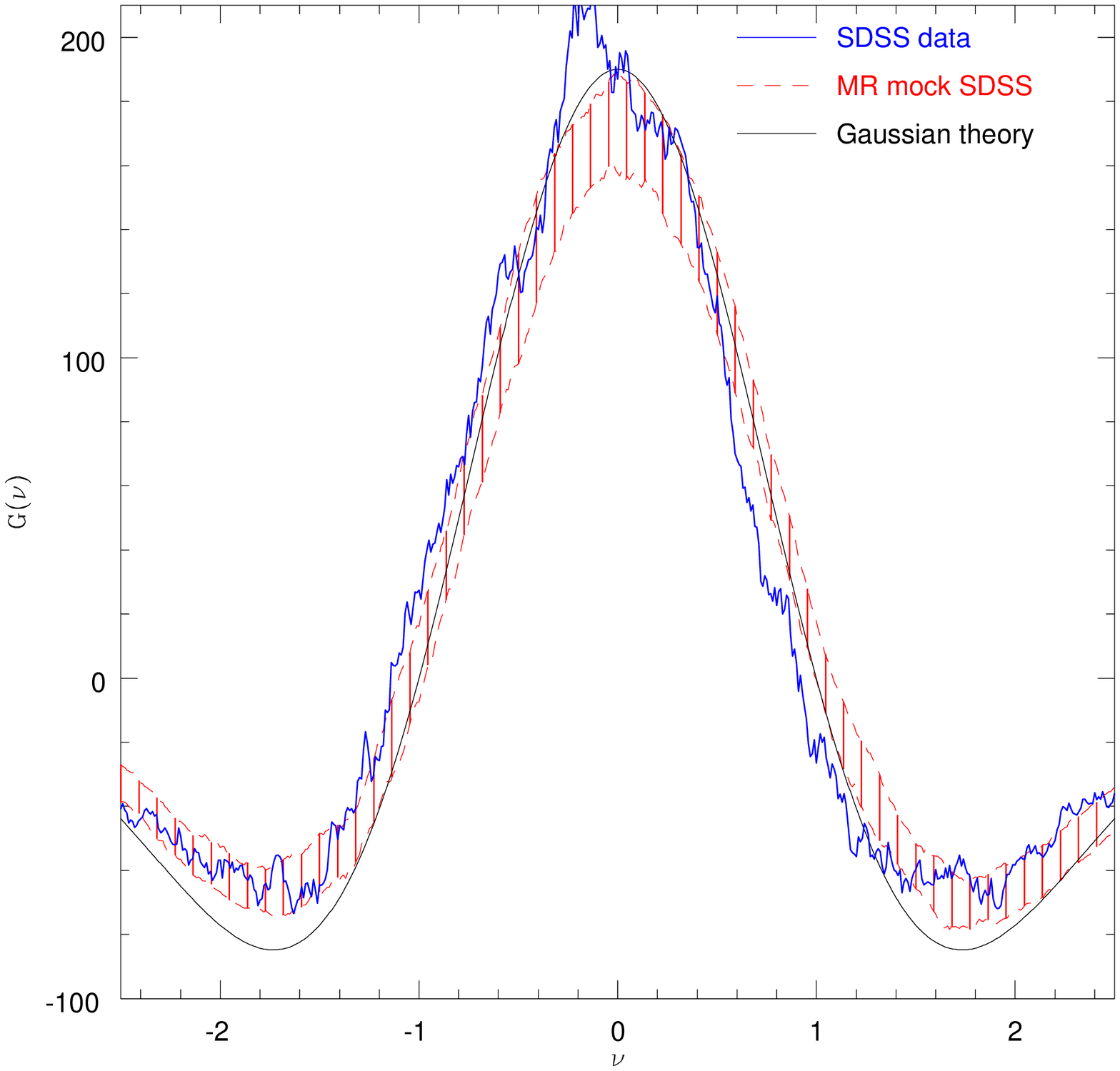}}}
\caption{Genus curves with shaded $1\sigma$
  error regions for the (a) 100 DH and (b) 50 MR samples, compared
  with SDSS and Gaussian random phase. }
\label{fig:genus:m}
\end{figure*}

Figure~\ref{fig:genus:r} shows the genus curve of the observations versus one
of the 50 MR mock surveys picked at random, one of
the 100 DH mock surveys picked at random, and the hydro mock
survey.  The DH simulation looks the best. The top of the
genus curve near $\nu  = 0$ is not cut off and it looks most like
the observations.  Both the Millennium Run mock and the hydro
mock run are cut off at the top in the same way, the hydro one
more so.  To get a better picture of the cosmic variance,
Figure~\ref{fig:genus:m} shows the observations compared to 
hatched bands showing the $1\sigma$ variation in the mock runs
from the (a) DH and (b) MR simulations.  It is clear
from this that the one random Millennium run mock survey shown
in figure~\ref{fig:genus:r} was worse than average at fitting the top of the
curve, but it is also clear that the DH simulation is still
better than the Millennium Run at fitting the observations
because the band of Millennium run simulations are lower in this
region.  The place where the Millennium Run mocks and the DH
mocks fail the worst is in the region  $0.4 < \nu< 1.2$ where the
observations are consistently shifted to the left with respect to the
simulations.  

We conclude that the simulations do an adequate job of representing
the topology in all the variables except $\Delta\nu$, which for the SDSS
data lies more than $2.5\sigma$ away from either the MR or DH
simulations.  
Of the 100 DH mock surveys
the one closest to the observations in terms of the four variables and
their error bars is mock catalog 94 which has $\Delta\nu = -0.02$, $A
= 191$, $A_v = 0.80$, $A_c = 0.77$ which are OK in all except
$\Delta\nu$ where it is still far off the observational value.  In
fact of the 100 mock DH simulations the two most negative in
$\Delta\nu$ are between $-0.05$ and $-0.04$ while the observations show
$-0.08$.  The four most negative of the 50 mock Millennium Run
simulations are between $-0.03$ and $-0.02$.  Interestingly, the higher
spatial resolution and semi-analytical modeling of the MR does not
yield a better fit to the observed topology of large-scale structure
than the dark-matter, halo-finding DH simulations.

\section{Conclusions}  

The SDSS dataset has now become large enough that the topology of
large-scale structure can be used for more detailed model testing.  We
find that the SDSS observations have a sponge-like median contour and
follow fairly closely the genus curve expected from Gaussian random
phase initial conditions predicted by inflation.  We quantify
departures from from this theoretical curve that provide key tests of
models for galaxy formation, as represented by the several simulations
that we examine.

The amplitude of the genus curve is in agreement with that predicted
from the standard $\Lambda CDM$ model (with the WMAP parameters) with
baryon oscillations included.  If baryon oscillations were not
included the fit to the amplitude would be significantly worse
($1.6\sigma$).  The observed values of $A_v$ and $A_c$ are predicted
well by both the MR and DH simulations. Both show the effects of
non-linear gravitational evolution and biased galaxy formation.  The
Cen and Ostriker hydro simulations are consistent with the data, but
their small volume gives them large error bars and they are currently
not giving values closer to the observations than the MR or
DH simulations.

The most notable feature of the observations is a 
meatball shift $\Delta\nu = -0.08$ showing a slight prominence of
isolated high density regions over isolated voids.  The SDSS
Great Wall is one large connected isolated high density region
and contributes to this effect, but the effect also shows up
in the northern part of the SDSS which does not
contain the Sloan Great Wall.  If the Sloan Great Wall were
entirely responsible for this result one might argue that it
was produced by rare objects whose frequency of occurrence was
determined by the power in the initial conditions at very
large scales and that even larger simulations $>(1024 h^{-1}
\mbox{Mpc})^3$ would be needed to properly test for this effect.  But 
this is not the case.  Negative values of $\Delta\nu = -0.08$ show up
even in the part of the survey that does not include the SDSS
Great Wall.  Also, slice surveys of the MR simulation show
great walls that look quite impressive---if not quite as
dramatic as  the Sloan
Great Wall. The observed $\Delta\nu$ values are more than $2.5\sigma$
away from the values found in the MR and the Dark
Matter Halo simulations. This is a severe test for large
N-body simulations and their heuristic galaxy formation
scenarios because these were not tuned to account for
topology.  

The slight meatball shift seen in the observations has been noticed in
previous observational samples with similar smoothing lengths ($6
h^{-1}$ Mpc), being mentioned first by Gott et al.\ (1989).  
The large survey by
Canaveses et al.\ (1998) of the IRAS galaxies looks Gaussian random
phase on all larger smoothing lengths, but does have a slight meatball
shift with a smoothing length of $5 h^{-1}$ Mpc.
Hydrodynamic simulations suggest that early type galaxies should show
more of a meatball shift than late type galaxies (Gott, Cen, \&
Ostriker 1996) and this effect has already been observed in the
equatorial slice of the SDSS in a 2D topology survey by comparing the
relative meatball shift between red and blue galaxies (Hoyle et al.\
2002).

As the SDSS Digital Sky Survey is completed, the gap between the
northern and equatorial slices will be closed giving us one large
continuous volume, where the fraction of the sample one throws away
because of closeness to the edge will be diminished.  This will
approximately double the effective volume of the sample and give us a
still better test. 
Also, studies with smoothing lengths of $10 h^{-1}$ Mpc and $20
h^{-1}$ Mpc will be possible with high precision allowing more direct
tests of the Gaussian random phase hypothesis on scales where the
galaxy formation effects are less important.  

It would be interesting to see N-body simulations covering larger
volumes which would have more power at large scales (because they would
not be artificially cut off at the box size).  This would make for more
accurate modeling of the structure and frequency of occurrence of
structures like the Sloan Great Wall \citep{gott05}.  

The results here suggest that in order to account for the observed
topology some changes in galaxy formation scenarios are called for.
We look forward to improvements in the N-body simulations.  Of
particular interest is how well larger hydrodynamic simulations will
perform when compared with larger samples, and whether there will be a
convergence of predictions as both hydrodynamic and merger tree, and
dynamical halo occupation methods are improved.  Galaxy formation is
not yet a solved problem in cosmology and the 3D topology offers a
strong test of models which is independent of other measures.

\section*{Acknowledgments}
This paper has been supported by JRG's NSF Grant AST04-06713. MSV
acknowledges support from NASA grant NAG-12243 and NSF grant
AST-0507463. MSV thanks the Department of Astrophysical Sciences at
Princeton University for its hospitality during sabbatical leave. MSV
and CBP thank the Aspen Center for Physics, at which some of this
work was completed. CBP acknowledges the support of the Korea Science
and Engineering 
Foundation (KOSEF) through the Astrophysical Research Center for the
Structure and Evolution of the Cosmos (ARCSEC),
and through the grant R01-2004-000-10520-0. RC acknowledges support
from NASA grant NNG05GK10G and NSF grant AST-0507521.

Funding for the SDSS and SDSS-II has been provided by the Alfred
P. Sloan Foundation, the Participating Institutions, the National
Science Foundation, the U.S. Department of Energy, the National
Aeronautics and Space Administration, the Japanese Monbukagakusho, the
Max Planck Society, and the Higher Education Funding Council for
England. The SDSS Web Site is \url{http://www.sdss.org/}.

The SDSS is managed by the Astrophysical Research Consortium for the
Participating Institutions. The Participating Institutions are the
American Museum of Natural History, Astrophysical Institute Potsdam,
University of Basel, Cambridge University, Case Western Reserve
University, University of Chicago, Drexel University, Fermilab, the
Institute for Advanced Study, the Japan Participation Group, Johns
Hopkins University, the Joint Institute for Nuclear Astrophysics, the
Kavli Institute for Particle Astrophysics and Cosmology, the Korean
Scientist Group, the Chinese Academy of Sciences (LAMOST), Los Alamos
National Laboratory, the Max-Planck-Institute for Astronomy (MPIA),
the Max-Planck-Institute for Astrophysics (MPA), New Mexico State
University, Ohio State University, University of Pittsburgh,
University of Portsmouth, Princeton University, the United States
Naval Observatory, and the University of Washington.

The Millennium Run simulation used in this paper was carried out by
the Virgo Supercomputing Consortium at the Computing Centre of the
Max-Planck Society in Garching. The semi-analytic galaxy catalogue is
publicly available at 
\url{http://www.mpa-garching.mpg.de/galform/agnpaper}, and additional
data are at \url{http://www.mpa-garching.mpg.de/millennium}.


\begin{thebibliography}{xxxxxxxxxxxxxxxxxxxxx}

\bibitem[Aarseth, Gott \& Turner (1979)]{agt79} Aarseth, S. J., Gott, J. R., \& Turner, E. L. 1979, ApJ, 228, 664
\bibitem[()]{abaz04} Abazajian, K., et al. 2004, \aj, 128, 502
\bibitem[()]{adel06}Adelman-McCarthy, J.K. et al. 2006, \apjs, 162, 38
\bibitem[()]{adler81} Adler, R. J. 1981, The Geometry of Random Fields
  (New York, Wiley)
\bibitem[()]{bert85} Bertschinger, E. 1985, ApJS, 58, 1
\bibitem[Bhavsar, Gott \& Aarseth (1981)]{bga81} Bhavsar, S. P., Gott, J. R., \& Aarseth, S. J. 1981, ApJ, 246, 656 
\bibitem[()]{blant03a}Blanton, M. R., Lin, H., Lupton, R. H., Maley, F. M., Young, N., Zehavi, I., \& Loveday, J. 2003a, \aj, 125, 2276 
\bibitem[()]{blant03b}Blanton, M. R., et al. 2003b, \aj, 125, 2348 
\bibitem[Blanton et al.(2005)]{blant05}Blanton, M. R., et al. 2005, \aj, 129, 2562 
\bibitem[Canavezes et al.(1998)]{cana98}Canavezes, A., et al. 1998, MNRAS, 297, 777
\bibitem[()]{cen94}Cen, R., Miralda-Escude, J., Ostriker, J. P., \& Rauch, M. 1994, ApJ, 437, L9 
\bibitem[Cen, Nagamine \& Ostriker(2005)]{cno05}Cen, R., Nagamine,
  K., \& Ostriker, J.P., 2005, \apj, 635, 86. 
\bibitem[()]{cenken}Cen, R., and Ostriker, K., 2006, in preparation
\bibitem[()]{}Choi, Y-Y, Park, C., \& Vogeley, M. S. 2006 \apj, in preparation
\bibitem[()]{}Colley, W. N. 1997, ApJ, 489, 471
\bibitem[()]{}Colley, W. N., \& Gott, J. R. 2003, MNRAS, 344 (3), 686
\bibitem[()]{}Colley, W. N., Gott, J. R., \& Park, C. 1996, MNRAS, 281, L82
\bibitem[()]{}Colley, W. N., Gott, J. R., Park, C., Weinberg, D., \& Berlind, A. 2000, ApJ, 529,795
\bibitem[()]{}Coles, P. 1988, MNRAS, 234, 509
\bibitem[Croton et al.(2006)]{croton06} Croton, D.J., et al. 2006, MNRAS, 365, 11
\bibitem[()]{}Davis, M., Efstathiou, G., Frenk, C.S., \& White, S. D. M. 1985, ApJ, 292, 371
\bibitem[()]{}de Bernardis, P., et al. 2000, Nature, 404, 955
\bibitem[()]{}de Lapparent, V., Geller, M., \& Huchra, J. 1986, ApJ, 302, L1
\bibitem[()]{}Doroshkevich, G. 1970, Astrophysika, 6, 320
\bibitem[()]{}Einasto, J., Joeveer, M., \& Saar, E., 1980, MNRAS,193,353
\bibitem[()]{}Eisenstein, D. J., et al. 2001, \aj, 122, 2267
\bibitem[()]{}Fillmore, J., \& Goldreich, P., 1984, ApJ, 281, 9
\bibitem[()]{}Fukugita, M., Ichikawa, T., Gunn, J. E., Doi, M., Shimasaku, K., \& Schneider, D. P. 1996, \aj, 111, 1748
\bibitem[()]{}Geller, M. J., \& Huchra, J. P. 1989, Science, 246, 897  
\bibitem[()]{}Gott, J. R., Cen, R. \& Ostriker, J. P. 1996, ApJ, 465, 499
\bibitem[Gott et al.(2006)]{gott06}Gott, J.R., Colley, W.N., Park, C-G., Park, C., Mugnolo, C. 2006, MNRAS (submitted)
\bibitem[Gott et al.(2005)]{gott05}Gott, J.R., Juric, M. et al., 2005, ApJ, 624 , 463
\bibitem[()]{}Gott, J. R., Mao, S., Park, C., \& Lahav, O. 1992, ApJ, 385, 26 
\bibitem[Gott, Melott, \& Dickinson (1986)]{gmd86} Gott, J. R., Melott, A. L. \& Dickinson, M. 1986, ApJ, 306, 341 
\bibitem[()]{}Gott, J. R. \& Rees, M. J. 1975, Astronomy and Astrophysics, 45, 365
\bibitem[()]{}Gott, J. R. \& Turner, E.L. 1977, ApJ, 216, 357
\bibitem[()]{}Gott, J. R. \& Turner, E.L. 1979, ApJ, 232, L79 
\bibitem[()]{}Gott, J. R., Turner, E.L., \& Aarseth, S. J. 1979, ApJ, 234, 13
\bibitem[()]{}Gott, J. R., Vogeley, M.S., Podariu, S. \& Ratra, B. 2001, ApJ, 549, 1
\bibitem[Gott, Weinberg \& Melott (1987)]{gwm87}Gott, J. R., Weinberg, D. N. \& Melott, A. L. 1987, ApJ, 319, 1 
\bibitem[()]{}Gott, J. R., et al. 1989, ApJ, 340, 625 
\bibitem[()]{}Gott, J. R., et al. 1990, ApJ, 352, 1
\bibitem[()]{}Gunn, J.E. \& Gott, J. R. 1972, ApJ, 176,1
\bibitem[()]{} Gunn, J.E., et al. 1998, \aj, 116, 3040
\bibitem[()]{}Gunn, J.E., et al. 2006, AJ, 131, 2332
\bibitem[()]{}Guth, A.H. 1981, Phys. Rev. D., 23, 247
\bibitem[()]{}Hamilton, A.J.S., Gott, J.R., \& Weinberg, D.W. 1986, ApJ, 309, 1
\bibitem[()]{}Hernquist, L. Katz, N., Weinberg, D.H., \& Miralda-Escude, J. 1996, ApJ, 457, L51 
\bibitem[()]{}Hikage, C., et al. 2002, Publ. Astron. Soc. Japan., 54, 707 
\bibitem[()]{}Hikage, C., et al. 2003, Publ. Astron. Soc. Japan, 55(5), 911 
\bibitem[()]{}Hogg, D. W., Finkbeiner, D. P., Schlegel, D. J., \& Gunn, J. E. 2001, \aj, 122, 2129
\bibitem[()]{}Hoyle, F., Vogeley, M.S., \& Gott, J. R. 2002, ApJ, 570, 44
\bibitem[()]{}Hoyle, F. et al.\ 2002, ApJ, 580, 663
\bibitem[()]{}Ivezic, Z. et al., AN, 325, 583
\bibitem[()]{}Kim, J., \& Park, C. 2006, ApJ, 639, 600
\bibitem[Kirshner et al.(1981)]{kirsh81}Kirshner, R. P., Oemler, A., Schechter, P. L., \& Shectman, S.A. 1981, ApJ, 248, L57 
\bibitem[()]{}Kogut, A.J., et al. 1996, ApJ, 464, L29
\bibitem[()]{}Linde, A.D. 1983, Physics Letters, 129B, 177
\bibitem[()]{}Lupton, R.H., Gunn, J.E., Ivezic, Z., Knapp, G.R., Kent, S., \& Yasuda, N. 2001, in ASP Conf. Ser. 238, Astronomical Data Analysis Software and Systems X, ed. F. R. Harnden, Jr., F. A. Primini, \& H. E. Payne (San Francisco: ASP), 269
\bibitem[()]{}Lynden-Bell, D. et al. 1988, ApJ, 326, 19L 
\bibitem[()]{}Matsubara, T. 1994, ApJ, 434, 43 
\bibitem[()]{}Melott, A. L., Cohen, A. P., Hamilton, A. J. S., Gott, J. R., Weinberg, D. H. 1989, ApJ, 345, 618 
\bibitem[()]{}Moore, B. et al., 1992, MNRAS, 256, 477.
\bibitem[()]{}Ostriker, J. P., \& Cowie, L. L. 1981, ApJ, 243, L127
\bibitem[()]{}Park, C. 1990, MNRAS, 242, 59P
\bibitem[Park(1991)]{parkthesis}Park, C. 1991, PhD Thesis, Princeton University
\bibitem[()]{}Park, C., Colley, W.N., Gott, J.R., Ratra, B., Spergel, D.N., \& Sugiyama, N. 1998, ApJ, 506,473 
\bibitem[()]{}Park, C., \& Gott, J.R. 1991a, MNRAS, 249, 288
\bibitem[()]{}Park, C., \& Gott, J. R. 1991b, \apj, 378, 457
\bibitem[()]{}Park, C., Gott, J.R. \& Choi, Y. J. 2001, ApJ, 553, 33 
\bibitem[()]{}Park, C., Gott, J.R., Melott,  A. \& Karachentsev, I. D. 1992, ApJ, 387,1
\bibitem[Park, Kim, \& Gott(2005)]{pkg05}Park, C., Kim,  J., \& Gott, J.R. 2005, ApJ, 633, 1
\bibitem[()]{park05}Park, C., et al.\ 2005, ApJ, 633, 11
\bibitem[Park et al.(2006)]{park06} Park, C., Choi, Y-Y, Vogeley, M.S., Gott,
  J.R., and Blanton, M.R. 2006, \apj, submitted
\bibitem[()]{}Park, C.-G. 2003, MNRAS, 349, 313
\bibitem[()]{}Pier, J. R., Munn, J. A., Hindsley, R. B., Hennessy, G. S., Kent, S. M., Lupton, R. H., \& Ivezic, R. 2003, \aj, 125, 1559
\bibitem[Peebles(1974)]{peebles74}Peebles, P. J. E. 1974, \apj, 189, L51
\bibitem[Peebles(1978)]{peebles78}Peebles, P. J. E. 1978, A\&A, 68, 345
\bibitem[()]{}Perlmutter, S., et al.\ 1999, ApJ, 517, 565
\bibitem[()]{}Press, W. H., \& Schecter, P. 1974, ApJ, 187, 425
\bibitem[()]{}Reiss, A. G., et al. 1998, AJ, 116, 1009
\bibitem[()]{}Rhoads, J. E., Gott, J. R., \& Postman, M. 1994, ApJ, 421, 1
\bibitem[()]{}Richards, G. T., et al. 2002, \aj, 123, 2945
\bibitem[()]{}Schlegel, D. J., Finkbeiner, D. P., \& Davis, M. 1998, \apj, 500, 525 
\bibitem[()]{}Shectman, S. A., et al. 1996, ApJ, 470, 172
\bibitem[()]{}Smith, J. A., et al. 2002, \aj, 123, 2121
\bibitem[()]{}Spergel, D. N., et al. 2003, ApJS, 148, 175
\bibitem[()]{}Spergel, D. N., et al. 2006, \apj, submitted, astro-ph/0603449
\bibitem[Springel et al.(2006)]{springel06}Springel, V., Frenk, C. S., \& White, S. D. M. 2006, Nature, 440, 1137
\bibitem[Springel et al.(2005)]{springel05}Springel, V., et al. 2005, Nature, 435, 629
\bibitem[()]{}Soneira, R. M., \& Peebles, P. J. E. 1978, AJ, 83, 845
\bibitem[()]{}Smoot, G. F., et al., 1994, ApJ, 437, 1
\bibitem[()]{}Stoughton, C., et al. 2002, \aj, 123, 485
\bibitem[()]{}Strauss, M. A., et al. 2002, \aj, 124, 1810
\bibitem[()]{}Tegmark,  M., et al., 2004, ApJ, 606, 702 
\bibitem[()]{}Tucker, D. et al. 2006, AN, in press
\bibitem[Vogeley et al.(1994)]{vogeley94} Vogeley, M. S., Park, C., Geller, M. J., Huchra, J. P., \& Gott, J. R. 1994, ApJ, 420, 525 
\bibitem[()]{} York, D., et al. 2000, \aj, 120, 1579 

\end{thebibliography}
\end{document}